\documentclass[prb, aps, amssymb, twocolumn, superscriptaddress, notitlepage, longbibliography]{revtex4-2}

\usepackage{amsmath}
\usepackage{graphicx}
\usepackage{color}
\usepackage[normalem]{ulem}
\usepackage[colorlinks]{hyperref}
\usepackage{dsfont}
\usepackage{physics}

\newcommand{\pd}{{\phantom{\dag}}}

\begin{document}

\title{Non-Hermitian
topology of transport in the quantum Hall phases in graphene}

\author{Raghav Chaturvedi}
\affiliation{Leibniz Institute for Solid State and Materials Research,
IFW Dresden, Helmholtzstrasse 20, 01069 Dresden, Germany}
\affiliation{W\"{u}rzburg-Dresden Cluster of Excellence ct.qmat, 01062 Dresden, Germany}

\author{Viktor K\"{o}nye}
\affiliation{Leibniz Institute for Solid State and Materials Research,
IFW Dresden, Helmholtzstrasse 20, 01069 Dresden, Germany}
\affiliation{W\"{u}rzburg-Dresden Cluster of Excellence ct.qmat, 01062 Dresden, Germany}
\affiliation{Institute for Theoretical Physics Amsterdam, University of Amsterdam,
Science Park 904, 1098 XH Amsterdam, The Netherlands}

\author{Ewelina M. Hankiewicz}
\affiliation{W\"{u}rzburg-Dresden Cluster of Excellence ct.qmat, 01062 Dresden, Germany}
\affiliation{Institute for Theoretical Physics and Astrophysics, Julius-Maximilians-Universit\"{a}t W\"{u}rzburg, D-97074 W\"{u}rzburg, Germany}

\author{Jeroen van den Brink}
\affiliation{Leibniz Institute for Solid State and Materials Research,
IFW Dresden, Helmholtzstrasse 20, 01069 Dresden, Germany}
\affiliation{W\"{u}rzburg-Dresden Cluster of Excellence ct.qmat, 01062 Dresden, Germany}
\affiliation{Department of Physics, TU Dresden, D-01062 Dresden, Germany}
\author{Ion Cosma Fulga}
\email{i.c.fulga@ifw-dresden.de}
\affiliation{Leibniz Institute for Solid State and Materials Research,
IFW Dresden, Helmholtzstrasse 20, 01069 Dresden, Germany}
\affiliation{W\"{u}rzburg-Dresden Cluster of Excellence ct.qmat, 01062 Dresden, Germany}

\date{\today}

\begin{abstract}
It has recently been shown that signatures of non-Hermitian topology can be realized in a conventional quantum Hall device connected to multiple current sources.
These signatures manifest as robust current-voltage characteristics, dictated by the presence of a nontrivial, non-Hermitian topological invariant of the conductance matrix. 
Chiral edge states are believed to be responsible for this non-Hermitian response, similar to how they lead to a quantized Hall conductivity in the presence of a single current source.
Here, we go beyond this paradigm, showing that multi-terminal conductance matrices can exhibit non-Hermitian topological phase transitions that cannot be traced back to the presence and directionality of a boundary-localized chiral mode.
By performing quantum transport simulations in the quantum Hall regime of monolayer graphene, we find that when the chemical potential is swept across the zeroth Landau level, unavoidable device imperfections cause the appearance of an additional non-Hermitian phase of the conductance matrix.
This highlights graphene as an ideal platform for the study of non-Hermitian topological phase transitions, and is a first step towards exploring how the geometry of quantum devices can be harnessed to produce robust, topologically-protected transport characteristics.
\end{abstract}

\maketitle

\section{Introduction} 
A physical system can have a non-Hermitian topological description if it is coupled to an environment and the resulting gains and losses are customized to reach a topological phase \cite{Ashida2020, Bergholtz2021}. 
Such a topological phase offers robust characteristics, with potential applications such as light funnels \cite{Weidemann2020}, amplifiers \cite{Wang2022}, and exponentially precise sensors \cite{Budich2020, könye2023nonhermitian}.
The experimental realization of these phases usually relies on platforms where gain and loss can be precisely controlled in each individual building block.
Examples include ultracold atoms \cite{Liang2022}, optical systems \cite{Xiao2020, Wang2021}, and meta-materials \cite{Helbig2020, Liu2021, Weidemann2020, Brandenbourger2019, Ghatak2020, Zhang2021a, Zhang2021b}. 

A recent study takes an alternative approach, which leverages the quantum transport properties of a well-known Hermitian topological phase -- the quantum Hall phase in a GaAs/AlGaAs heterostructure \cite{Ochkan2024}. It is based on the insight that in the limits of maximum non-reciprocity, the Hamiltonian for a Hatano-Nelson (HN) chain \cite{Hatano1996} accurately describes the long-time dynamics of the chiral edge states in a quantum Hall system \cite{Lee2019}. 

In the presence of multiple current sources, the current-voltage characteristics of the quantum Hall phase show robust features that originate from the non-Hermitian topology of the conductance matrix, not of the Hamiltonian.
The chiral edge modes ensure that charge propagates between neighboring contacts only in one direction, meaning that the multi-terminal conductance matrix becomes numerically equivalent to the Hamiltonian matrix of the HN chain, a prototypical model of non-Hermitian topology. In effect, each terminal plays the role of a `site' in the HN chain, while the edge modes propagating between adjacent contacts play the role of nearest-neighbor, one-way `hoppings.' 

Away from the quantum Hall plateau regime, the bulk becomes conducting and diffusive transport occurs between any pair of contacts, such that the analogy between chiral edge modes and nonreciprocal nearest-neighbor hoppings breaks down.
All elements of the non-Hermitian conductance matrix become sizeable, potentially leading to additional topological phases and phase transitions that are not the direct consequence of the chiral modes.
Here, we demonstrate the emergence of an additional phase in the non-Hermitian, multi-terminal conductance matrix of Chern insulators. 
This extra phase is not caused by boundary-localized chiral modes, arising instead in the transition region between two quantum Hall plateaus with opposite Chern numbers.
The topologically trivial phase is due to the inhomogeneities present in the conductance matrix, which occur in any experimental setup. 
Our work shows that, beyond the quantum Hall regime, the quality and geometry of multi-terminal devices can play an important role in generating robust current-voltage characteristics protected by non-Hermitian topology.

To illustrate this concept, we turn to graphene, one of the most accessible and highly-tunable platforms for realizing quantum Hall physics. In the GaAs/AlGaAs heterostructure based 2DEG used in the previous study \cite{Ochkan2024}, the direction of edge state propagation could be reversed by reversing the direction of the magnetic field, which would involve going through a classical Hall regime. A graphene based experimental setup would offer more tunability, as the direction of the edge states can additionally be reversed through the application of a gate voltage.

Using a toy model, we first highlight the correspondence between chiral edge modes and the non-Hermitian topology of the conductance matrix. 
This correspondence remains valid even as the chemical potential is swept through the charge neutrality point and the edge modes change direction.
Next, we introduce inhomogeneities into the multi-terminal conductance matrix, showing that they cause the emergence of an additional phase. This additional non-Hermitian trivial phase corresponds to the vanishing of the exponential profile of current and voltage vectors, and this occurs in a regime where graphene has finite non-reciprocal conductance between its terminals, i.e., the non-Hermitian trivial phase of the conductance matrix cannot be traced back to a trivial insulating phase in graphene.

\section{Transport in graphene and its connection to non-Hermitian topology ---} 
\begin{figure}[tbh!]
\centering
\includegraphics[width = 0.45\textwidth]{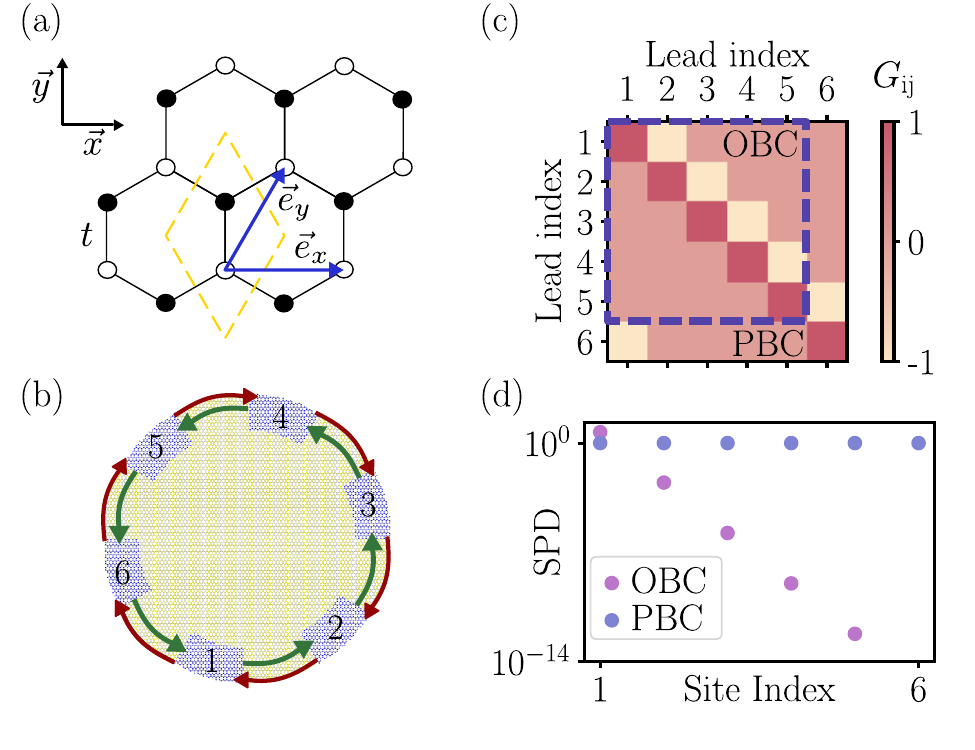}
\caption{
Panel (a): Graphene lattice with two sites (shown as white and black circles) in each unit cell (marked by the yellow dashed contour). 
$\Vec{e}_x^{\pd}$ and $\Vec{e}_y^{\pd}$ are the Bravais vectors and $t$ is the nearest neighbor hopping. Panel (b): Schematic of a graphene disk with six contacts (blue) attached to its boundaries and labeled from one to six. 
The transmission between adjacent leads in the clockwise and anticlockwise directions is represented by red and green arrows, respectively. 
Panel (c): Conductance matrix $G$ for a disk of graphene (radius $r = 40$ in units of the lattice constant) with six leads [as in (b)], deep in a quantum Hall phase with Chern number $-1$ ($\mu/t = -0.2$, $\phi=0.18$). 
The conductance (in units of $e^2/h$) is quantized between adjacent leads in the downstream direction of the edge modes and is zero in the opposite direction. 
The full conductance matrix is equivalent to a Hatano-Nelson Hamiltonian under PBC whereas the sub-matrix (enclosed by blue dashed lines) is equivalent to the same under OBC. 
Panel (d): SPD plotted for the $G$ matrix in panel (c) against the site index (lead index) on a log-linear scale under OBC (non-Hermitian skin effect) and PBC.
\label{fig:Main:Schematics}
}
\end{figure}

We use a toy model for graphene consisting of spinless fermions hopping on a honeycomb lattice. The Hamiltonian takes the form 
\begin{equation} \label{eq-graphene}
    H = t \sum_{\langle ij \rangle}^{\pd} c^{\dag}_i c^{\pd}_j + \mu \sum_i c_i^{\dag} c^{\pd}_i,
\end{equation}
where $\mu$ represents the on-site energy of an electron at the site index $i$, $c^{\pd}_{i}$ $(c^\dag_{i})$ represent its annihilation (creation) operator, $\langle\dots\rangle$ denotes nearest neighbors, and $t$ is the hopping parameter. 
The unit cells are labeled by integers $(n_x^{\pd},n_y^{\pd})$ in the $\Vec{e}_x^{\pd}, \Vec{e}_y^{\pd}$ directions,  respectively [see Fig.~\ref{fig:Main:Schematics}(a)]. 

We introduce a perpendicular magnetic field by changing the vertical hoppings in Fig.~\ref{fig:Main:Schematics}(a) from $t$ to $t \exp(i\phi n_x^{\pd})$, where $\phi/2\pi$ is the flux per hexagon, in units of the flux quantum. 
This gaps out the Dirac cones and results in the formation of Landau levels, symmetrically distributed around the charge neutrality point, $E=0$ \cite{Novoselov2005} (see Appendix \ref{SM:sec:band_structure} for details).
Since the model is spinless, sweeping the chemical potential from $\mu>0$ to $\mu<0$ reverses the edge mode propagation direction from counter-clockwise to clockwise, leading to a direct transition between phases with Chern numbers $+1$ and $-1$.

Using the Kwant package \cite{Groth2014}, we attach $N$ equally-spaced ideal leads to the boundary of a finite-sized, disk-shaped system, as shown in Fig.~\ref{fig:Main:Schematics}(b) for $N=6$. Numerical details are described in Appendix \ref{SM:sec:numerical_detials}.
This models an experimental setup in which $N$ contacts are connected to a graphene sample.
In the linear regime, the currents injected into each of the contacts, $\vec{I} = (I_1, \ldots, I_N)$, are related to the potentials on each of the contacts, $\vec{V} = (V_1, \ldots, V_N)$, as $\vec{I} = G \vec{V}$, where $G$ is the $N\times N$ conductance matrix.
At zero-temperature and in the zero bias voltage limit, the elements of $G$ take the form
\begin{equation}\label{eq:Gij}
    G_{ij} = \frac{e^2}{h} \left[ \delta_{ij} {\cal N}_j - {\rm tr} \left( S^\pd_{ij} S^\dag_{ij} \right) \right].
\end{equation}
Here, $e$ is the electric charge, $h$ is Planck's constant, $\delta_{ij}$ denotes the Kronecker delta, ${\cal N}_j$ is the number of propagating modes in lead $j$, and ${\rm tr}$ denotes the trace. The matrix $S_{ij}$ is a block of the full, multi-terminal scattering matrix, containing the probability amplitudes for a particle to be transmitted from any incoming mode of lead $j$ to any outgoing mode of lead $i$.

For the six-terminal system shown in Fig.~\ref{fig:Main:Schematics}(b), setting $\mu=-0.2 t$ and $\phi=0.18$, such that the system is deep in the quantum Hall phase with Chern number $-1$, leads to the conductance matrix shown in Fig.~\ref{fig:Main:Schematics}(c).
This matrix is identical to the non-Hermitian Hamiltonian matrix of the Hatano-Nelson tight-binding model \cite{Hatano1996}, a chain of orbitals connected by nonreciprocal nearest-neighbor hoppings. 
Thus, each of the diagonal elements $G_{ii}$ can be thought of as being analogous to the onsite energy of a site in the fictitious chain. 
The off-diagonal elements are analogous to hopping integrals connecting the different fictitious sites.
In this case, each of the sites has an onsite energy equal to $+1$, one-way hoppings equal to $-1$ connect the $j^{\rm th}$ and $(j-1)^{\rm th}$ sites, and since $G_{61}=-1$, the chain has periodic boundary conditions (PBC).
Realizing an analogue of the Hatano-Nelson model with open boundary conditions (OBC) can be done by grounding the $6^{\rm th}$ contact (setting the potential $V_6=0$).
In this case the currents and voltages corresponding to the first five terminals would be related by the $5\times 5$ conductance sub-matrix, $G_{\rm OBC}$, shown by the dashed box in Fig.~\ref{fig:Main:Schematics}(c).

The nontrivial topology of the Hatano-Nelson Hamiltonian matrix, and thus of the multi-terminal conductance matrix, results in robust current-voltage characteristics.
A transport regime in which the current vector is proportional to the voltage vector $\vec{I}\propto\vec{V}$ can only be reached provided that these vectors are eigenvectors of the conductance matrix.
In the OBC case (grounded $6^{\rm th}$ contact), all eigenvectors are exponentially localized to one end of the fictitious Hatano-Nelson chain, a non-Hermitian topological phenomenon referred to as the non-Hermitian skin effect \cite{Alvarez2018, Yao2018}.
This can be seen from the probability density summed over all right eigenvectors $v_n$ of the conductance matrix, ${\rm SPD}=\sum_n |v_n|^2$.
Thus, when $\vec{I}\propto\vec{V}$ the currents and voltages on terminals $1$ to $5$ follow the exponentially decaying pattern shown in Fig.~\ref{fig:Main:Schematics}(d).

In contrast, for the PBC setup, when $\vec{I}\propto\vec{V}$ the six currents and voltages do not show an exponential profile.
Both of these proportionality regimes can be reached in a robust and predictable way by means of iterative transport measurements, as shown experimentally in Ref.~\cite{Ochkan2024}.

\section{Inhomogeneity and topological phase transitions ---} 
We are interested in studying the topology of the conductance matrix away from the quantum Hall plateau regime, when bulk transport is significant and all elements of $G$ are nonzero.
To this end, we follow Refs.~\cite{Hughes2021, Ochkan2024}, and focus only on the OBC setup in the following.
The $N^{\rm th}$ contact is grounded at all times and only the $(N-1)\times(N-1)$ sub-block of the conductance matrix, $G_{\rm OBC}$, is considered.
We use a topological invariant, $w_{\rm PD}^{\pd}$, that can be determined directly from this conductance matrix, and which predicts the existence of a non-Hermitian skin effect.
To compute it, we first subtract the average of the diagonal elements from the conductance matrix, $\hat{G} = G_{\rm OBC} - \lambda \mathds{1}$, where $\lambda = {\rm tr}( G_{\rm OBC}) / (N-1)$ and $\mathds{1}$ is the identity matrix.
Writing the polar decomposition of this matrix, $\hat{G} = QP$, where $Q$ is unitary and $P$ is a positive-definite matrix, yields

\begin{equation} \label{eq:wpd_invariant}
    w_{\rm PD}^{\pd}(\hat{G}) = \mathcal{T}(Q^{\dag}[Q, X]),
\end{equation}
where $X={\rm diag}(1, 2, \ldots, N-1)$ encodes the lead index, and $\mathcal{T}$ is the trace per unit volume evaluated over the middle indices, far from the ends (see Appendix \ref{SM:sec:numerical_detials} for details). 

\begin{figure}[tb]
\centering
\includegraphics[width= 0.45\textwidth]{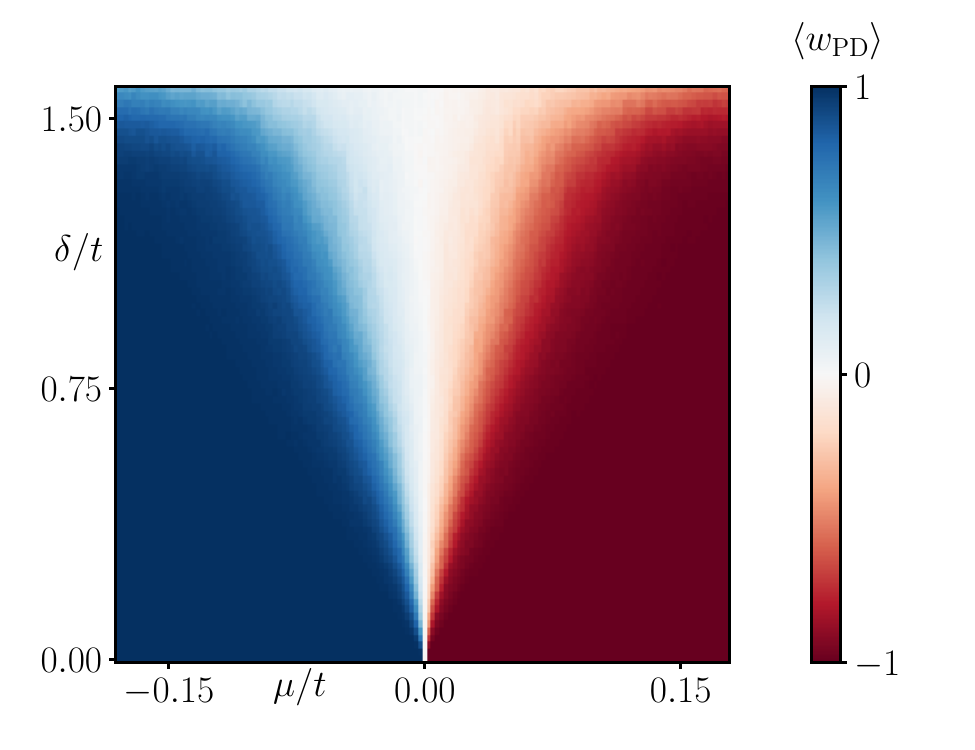}
\caption{
$\langle w_{\rm PD}^{\pd} \rangle$ evaluated for conductance matrices of a monolayer spinless graphene disk ($r = 60$) with 10 leads, one of which is grounded (9 lead OBC matrix) as a function of varying $\delta/t$ and chemical potential $\mu/t$. 
The trivial phase ($\langle w_{\rm PD}^{\pd} \rangle = 0$) becomes more prominent close to the charge neutrality point $\mu/t = 0$ as $\delta$ increases. Here, $\phi = 0.18$. \label{fig:Main:Graphene_phase_diagram}
}
\end{figure}

As expected, we confirm that $w_{\rm PD}^{\pd}$ correctly predicts the appearance of a non-Hermitian skin effect in the conductance matrix of Fig.~\ref{fig:Main:Schematics}(d).
Deep in the quantum Hall plateau regime, where the boundary-localized chiral modes produce one-way nearest-neighbor hoppings of the fictitious Hatano-Nelson chain, the invariant in Eq.~\eqref{eq:wpd_invariant} is locked to the Chern number.
We have verified that sweeping the chemical potential across the charge neutrality point leads to a direct transition in the Chern number from $+1$ to $-1$, and simultaneously to a direct transition of $w_{\rm PD}^{\pd}$, from $-1$ to $+1$. 
Correspondingly, there is a change of direction in the non-Hermitian skin effect: all eigenvectors of the conductance matrix are all localized either on the left or the right side of the chain, depending on the sign of $\mu$.

In any experiment, we expect that the conductance matrix will be inhomogeneous due to multiple factors.
These include the varying quality of different contacts, unequal distances between them, disorder and defects in the sample, as well as sample-specific quantum fluctuation effects (universal conductance fluctuations) \cite{Beenakker1997}.

For simplicity, we obtain inhomogeneous conductance matrices $G$ by randomly changing the potential on each site of the honeycomb lattice, $\mu \rightarrow \mu + \delta_i^{\pd}$, where $\delta_i^{\pd}$ is chosen from a uniform distribution $(-\delta,\delta)$. 

The average non-Hermitian topological invariant of the conductance matrix can be expressed as:
\begin{equation} \label{eq:wpd-method01}
    \langle w_{\rm PD}^{\pd} \rangle = \frac{\sum_{i=1}^{n} w_{\rm PD}^{\pd}(\hat{G_i})}{n},
\end{equation}
for $n$ independent realizations of the random onsite potential, each producing a different conductance matrix $G_i$.

\begin{figure}[tb]
\centering
\includegraphics[width=0.45\textwidth]{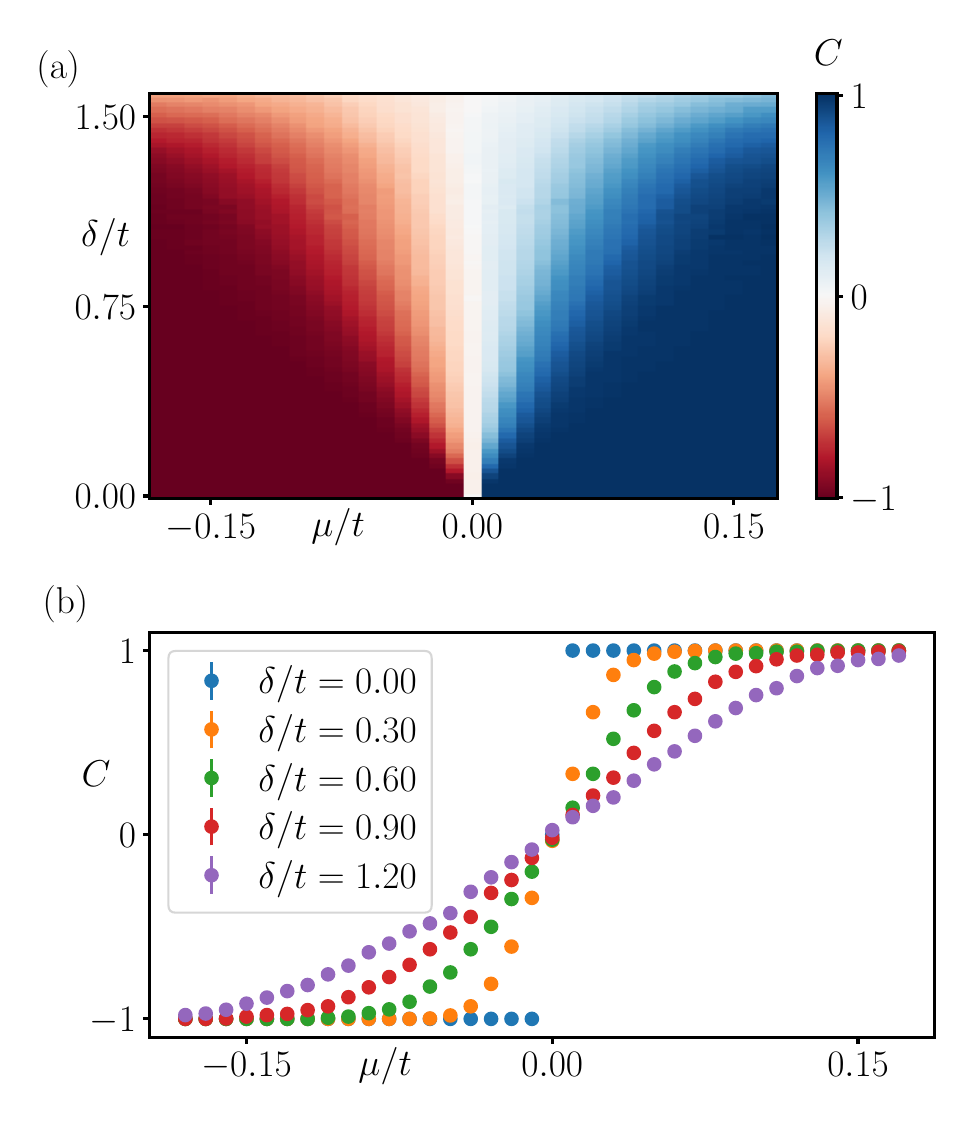}
\caption{ Panel (a): Disorder-averaged Chern number ($C$) for a monolayer spinless graphene disk (r=15) plotted as a function of the disorder strength $\delta/t$ and chemical potential $\mu/t$. Panel (b): Horizontal line cuts from panel (a) showing the disorder-averaged Chern number exhibiting a direct transition, i.e. varying continuously between $\pm 1$ for various non-zero disorder strengths $\delta/t$, without any intermediate trivial phase. Here, $\phi = 0.18, \kappa = 0.01$.
\label{fig:Main:Graphene_Chern_number}
}
\end{figure}

\begin{figure}[tb]
\centering
\includegraphics[width=0.45\textwidth]{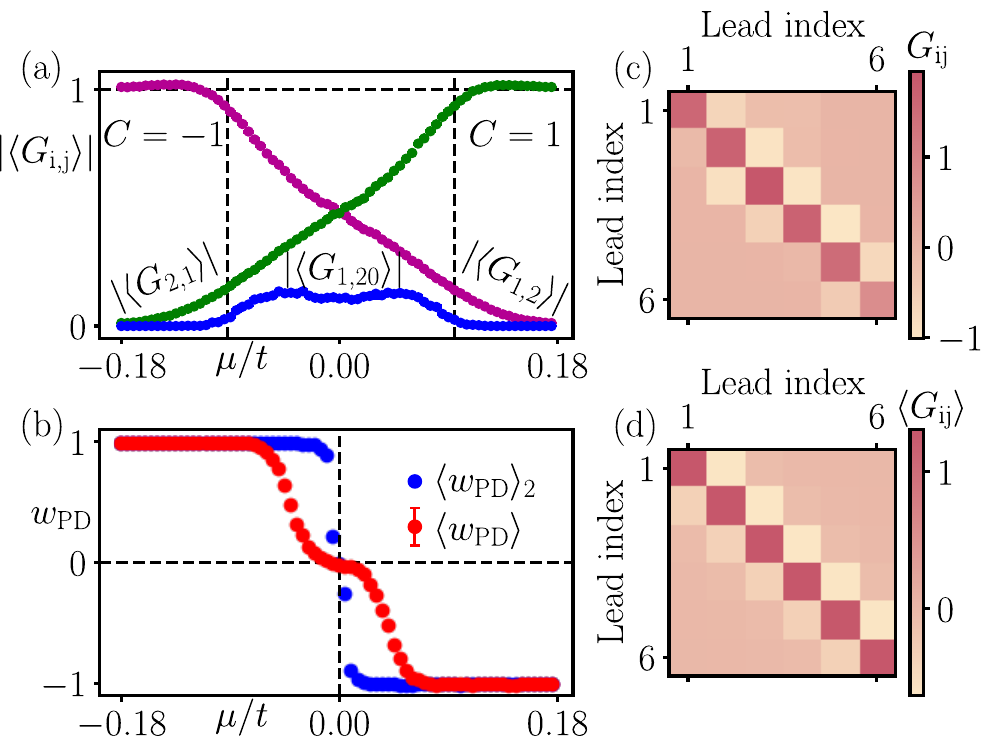}
\caption{ Panel (a): Absolute value of disorder-averaged conductances $\langle G \rangle_{1,2}$ (purple) and $\langle G \rangle_{2,1}$ (green), showing the transition between two quantum Hall phases with Chern numbers $C=1$ and $C =-1$, for a spinless graphene disk ($r = 240$) with $N=40$ contacts. In these phases, the conductance through the bulk is negligible, as indicated by the $\langle G \rangle_{1,20} \times 100$ (multiplied by a factor of 100 for ease of visibility) (blue). This bulk conductance becomes non-zero in the transition region between the two quantum Hall phases, indicating the diffusive transport regime [see Appendix \ref{SM:sec:Longer_range_couplings}].    
Panel (b): evaluated $\langle w_{\rm PD}^{\pd} \rangle$ shows the presence of a trivial phase close to $\mu/t = 0$ between two non-Hermitian topological phases ($+1$ and $-1$). This is in contrast to $\langle w_{\rm PD}^{\pd} \rangle_2$ which shows a direct transition between the two phases. Here, $\delta/t = 1.2$, and $\phi = 0.18$. Panel (c): A $6 \times 6$ conductance sub-matrix at $\mu = -0.04t$ for a single realization of onsite disorder, indicating the inhomogeneities in its entries. Panel (d): A $6 \times 6$ sub-matrix of the disorder-averaged conductance matrix $\langle G \rangle $ at $\mu = -0.04t$ exhibiting the uniformity in its entries (see Appendix \ref{SM:sec:wpd_scaling}).
\label{fig:Main:Graphene_phases_disc}}
\end{figure}

Figure \ref{fig:Main:Graphene_phase_diagram} shows that for $\delta=0$ the conductance matrix exhibits a direct transition from $w_{\rm PD}^{\pd}= +1$ to $-1$ as the chemical potential is swept across the charge neutrality point, as mentioned above. 
Turning on inhomogeneities, however, leads to the appearance of a trivial phase of the conductance matrix, $\langle w_{\rm PD}^{\pd} \rangle=0$, for which the current-voltage characteristics do not show the exponential profile of the non-Hermitian skin effect (see Appendix \ref{SM:sec:wpd_scaling}).
We observe that increasing $\delta$ leads to an enlargement of this phase, which appears for a wider range of chemical potential $\mu$.

The intervening trivial non-Hermitian phase of the conductance matrix appears in a region where the graphene disk is not a trivial insulator. We confirm this by computing a local Chern marker for the graphene disk, known as the Localizer index. This is a topological invariant evaluated using real-space methods and is equivalent to the Chern number $C$ for quantum Hall systems \cite{Loring2015, Loring2019}. In particular, using the position operators $\Bar{X} $ and $\Bar{Y}$ of the graphene disk encoding the position information of the lattice sites and the Hamiltonian $H$, a composite Hermitian operator known as the spectral Localizer $L$ is constructed as:
\begin{equation} \label{eq:Localizer}
    L (\Bar{X},\Bar{Y}, H) = \kappa[\sigma_x \otimes \Bar{X} + \sigma_y \otimes \Bar{Y}] + \sigma_z \otimes H,
\end{equation}
where $\kappa$ is a scaling factor chosen appropriately to ensure compatible weights for $H, \Bar{X}, $ and $\Bar{Y}$ \cite{Loring2015, Loring2019}. The topological index is then computed as:
\begin{equation} \label{eq:Localizer_index}
    C = \frac{1}{2} \text{\rm sig} [L(\Bar{X}, \Bar{Y},H)],
\end{equation}
where sig refers to the signature of a matrix (number of positive eigenvalues minus the number of negative eigenvalues). Fig.~\ref{fig:Main:Graphene_Chern_number} shows that for non-zero disorder strength $\delta$, the topological index (equivalent to the Chern number $C$) exhibits the absence of a trivial insulating phase with zero Chern number, i.e. it varies continuously between the values $\pm 1$ without having a plateau at zero.

\begin{figure}[tbh] 
\centering
\includegraphics[width=0.45\textwidth]{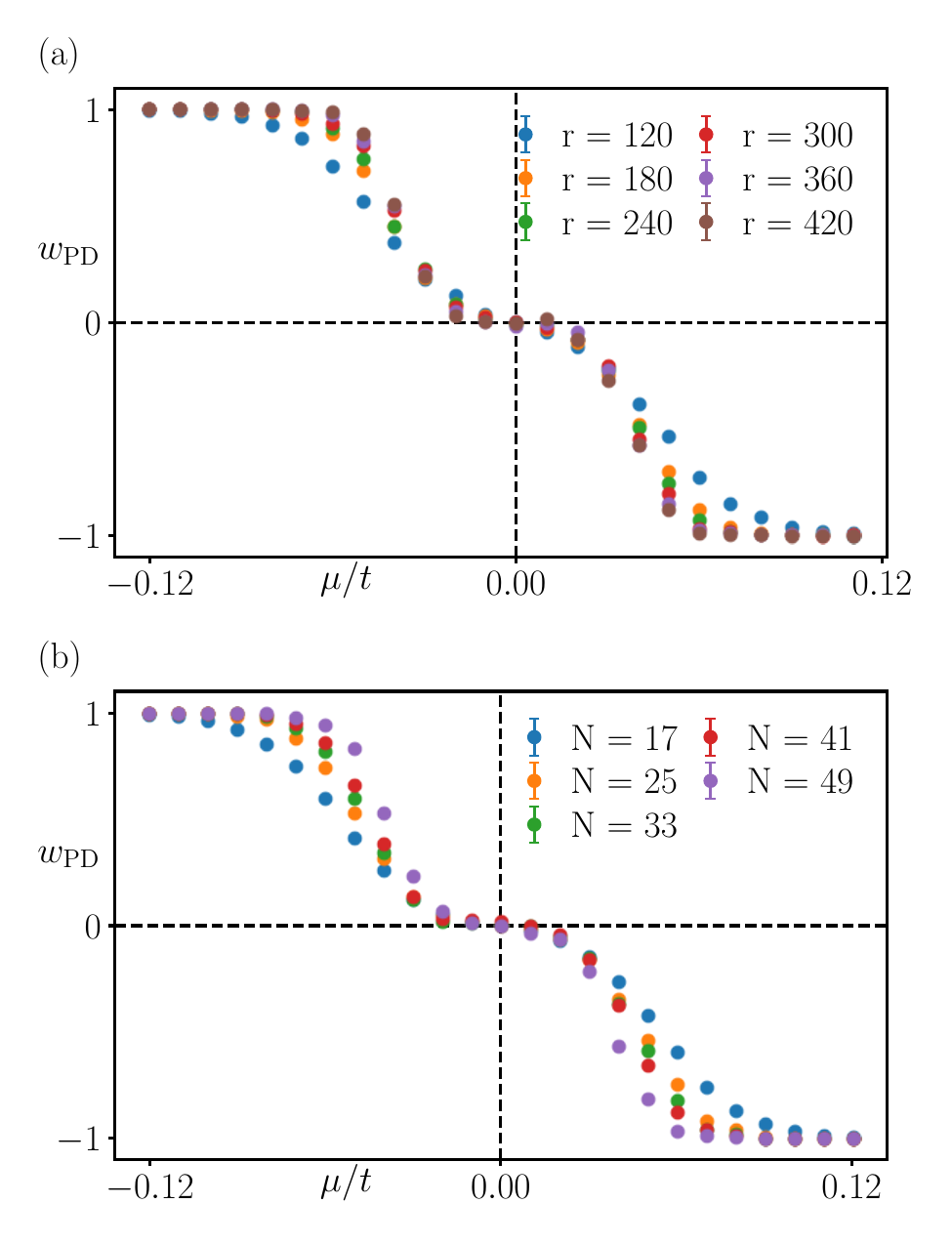}
\caption{Finite-size scaling: (a) $\langle w_{\rm PD}^{\pd} \rangle$ for a disk with increasing radius $r$ and proportionately increasing number of leads $N = r/6$ showing that the trivial phase gets better resolved as the size of the graphene disk increases. (b) $\langle w_{\rm PD}^{\pd} \rangle$ for different sizes $N$ of the OBC conductance matrix $G_{\rm OBC}^{\pd}$ as a function of the chemical potential $\mu$ indicating the presence of the trivial phase for different number of leads. Here, $\delta = 1.2t$.}\label{fig:Main:wpd_scaling}
\end{figure}

In the region where the non-Hermitian trivial phase develops, the bulk of the graphene is conducting, due to the fact that random potential fluctuations broaden the zeroth Landau level. This is shown in Fig.~\ref{fig:Main:Graphene_phases_disc} (a) as a non-zero conductance value across the graphene disk  $G_{1,20}$ for a disk with 40 contacts and also in Appendix \ref{SM:sec:Longer_range_couplings}. In the thermodynamic limit, the transport through the bulk will vanish for all chemical potentials except at the quantum Hall plateau transition $\mu = 0$. On the level of the effective Hatano-Nelson chain realized by the multi-terminal conductance matrix, the conductance between adjacent leads mimics the  ``hopping'' between adjacent ``sites'' of the chain. Thus, in order to meaningfully go towards the thermodynamic limit of the effective Hatano-Nelson chain we have to scale up both the radius of the graphene disk as well as the total number of contacts, in such a way that the distance between neighboring contacts remains constant. We implement this finite-size scaling study by increasing the radius $r$ of the graphene disk while simultaneously increasing the number of contacts $N$ attached to its boundaries ($N = r/6$), grounding 1 lead every time to get the OBC matrix. The corresponding $\langle w_{\rm PD}^{\pd} \rangle$ averaged over 1000 disordered configurations shows that the extra non-Hermitian phase gets better resolved as the size of the original graphene disk increases (see Fig.~\ref{fig:Main:wpd_scaling}(a)). In another scaling study, we fix the size of the graphene disk ($r = 300$) and change the number of contacts used in the calculation of the conductance matrix. We start by attaching 50 leads to the graphene disk and ground $p$ out of these to give a $(50-p) \times (50-p)$ conductance matrix under OBC. The scaling behavior of the $\langle w_{\rm PD}^{\pd} \rangle$ (averaged over 1500 independent onsite disorder realizations) is shown in Fig.~\ref{fig:Main:wpd_scaling}(b) for different choices of $p$ corresponding to different sizes $N$ of the OBC conductance matrix, again indicating that the trivial phase is better resolved as $N$ increases. Finally, we emphasize that the equivalence between the topology of a d-dimensional system and its (d-1) dimensional edge \cite{Lee2019} could be obtained in our setting when the conductance matrix has contributions solely from the edge. This would correspond to the case where the radius of the graphene disk is made infinitely large without increasing the number of contacts. The separation between the contacts in this case will be much larger than the localization length of the bulk states, resulting in a complete suppression of bulk transmission.


The presence of conducting bulk modes is not the cause of the the non-Hermitian trivial phase. To show this, we consider a different averaging scheme, where we first average the conductance matrix over the random realizations of onsite disorder, and only afterwards compute the topological invariant:
\begin{equation} \label{eq:wpd-method02}
    \langle w_{\rm PD}^{\pd} \rangle_{2}^{\pd} = w_{\rm PD}^{\pd}\left(\frac{1}{n}\sum_{i=1}^{n}\hat{G_i}\right).
\end{equation}
This procedure leads to a conductance matrix with uniform entries on each diagonal, where the broadening of the zeroth Landau level is still present, but does not produce an extra non-Hermitian phase in the conductance matrix (see Fig.~\ref{fig:Main:Graphene_phases_disc}). 

The true reason for emergence of the extra phase between 
$\langle w_{\rm PD}^{\pd} \rangle = \pm 1$ phases is due to inhomogeneities in the conductance matrix and is a natural consequence of the current conservation rules. 
Since the sum of all elements of $G$ over any column must vanish due to Kirchhoff's laws, $\sum_{i = 1}^L G_{ij}^{\pd} = 0$, random conductance fluctuations between different pairs of leads necessarily lead to random changes on the conductance matrix diagonal.
On the level of the fictitious, 1D Hatano-Nelson chain simulated by $G$, this corresponds to the presence of onsite disorder, which is known to lead to an extra $\langle w_{\rm PD}^{\pd} \rangle=0$ phase (see Ref.~\cite{Hughes2021} and Appendix \ref{SM:sec:HN_trivial_phase}).

\section{Conclusion} 
We have theoretically investigated the non-Hermitian topological behavior associated with multiple-current-source transport in Chern insulators, using the quantum Hall phases of a honeycomb graphene model as an example. 
Our simulations demonstrated the presence and robustness of non-Hermitian phases in the conductance matrix of this system, which can easily be tuned through chemical potential variations. 
This variation can be achieved in practice via gate voltage adjustments, making graphene a promising platform to realize current-voltage relations protected non-Hermitian topology. 

Interestingly, we found that inhomogeneities in the conductance matrix, which can arise, e.g. due to varying contact quality or inter-contact distance lead to the emergence of an additional, trivial non-Hermitian phase in the system's conductance matrix.
This implies that away from the quantum Hall plateau regime, the quality and geometry of quantum devices may play a decisive role in producing transport properties associated with non-Hermitian topology. 
Further, while we have focused on a graphene toy model for simplicity, we expect this behavior to be generic to a larger class of Chern insulators, including those realized in magnetically doped topological insulators \cite{Chang2013}. 

\section{Data availability}
The data and code used to generate the figures are available at \cite{zenodocode}.

\section{Acknowledgments} 
We thank Ulrike Nitzsche for her technical assistance. 
We acknowledge funding by the Deutsche Forschungsgemeinschaft (DFG, German Research Foundation) through SFB 1170, Project-ID 258499086, and through the W\"urzburg-Dresden Cluster of Excellence on Complexity and Topology in Quantum Matter – ct.qmat (EXC2147, Project-ID 390858490). V.K. was funded by the European Union.

\bibliography{main.bib}

\begin{appendix}

\section{Bandstructure} \label{SM:sec:band_structure}

\begin{figure}[tbh]
\centering
\includegraphics[width=0.45\textwidth]{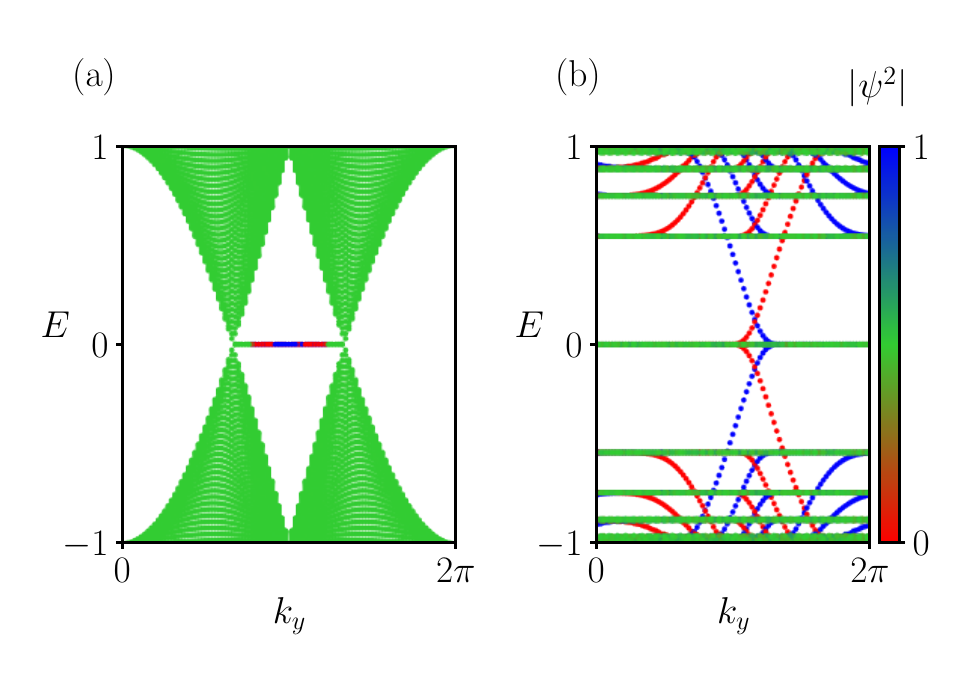}
\caption{
Bandstructure of a ribbon of the graphene toy model with translation invariance along the $\Vec{e_y}$ direction ($\mu/t = 0$).  
Panel (a), $\phi = 0$: Dirac cone spectra in the absence of a magnetic field.
Panel (b), $\phi = 0.18$: Landau levels emerge in the presence of a magnetic field, alongside unidirectional propagating boundary-localized modes at the ribbon's edges. The color scale represents the probability density of energy eigenstates integrated over half of the total number of unit cells in the finite direction $(0 \leq n_x \leq L/2)$. Eigenstates localized at opposite boundaries are colored in red and blue, respectively. 
\label{fig:SI:Bandstructure}
}
\end{figure}

In this section, we present the energy band diagrams for the graphene toy model described in the main text. We consider the system in a ribbon geometry with translational invariance along the $\Vec{e_y}$ direction. We fix $L = 100$ unit cells in the finite width direction $\Vec{e_x}$, 

In the absence of a magnetic flux through the system, the spectrum has a gapless Dirac cone structure with two Dirac cones connected by zig-zag edge modes [see Fig.~\ref{fig:SI:Bandstructure}(a)]. 

In the presence of a magnetic field, a gap opens in this spectrum and Landau levels emerge in the bulk [see Fig.~\ref{fig:SI:Bandstructure}(b)] alongside edge localized propagating boundary modes. Here, a Landau level exists at the charge neutrality point ($E = 0$) which separates two regions of opposite Chern numbers. 

\section{Details of numerical simulations} \label{SM:sec:numerical_detials}

This section provides details of the numerical simulations for the results in the main text. The graphene toy model is described on a honeycomb lattice using the Kwant package  \cite{Groth2014}. Here, the lattice constant is equal to 1, and all lengths are measured in its units.

For Fig.~\ref{fig:Main:Schematics},  the radius of the finite-sized disk is fixed at $r = 40$. We attach 6 equally spaced ideal leads to the sites of the system close to the disk's outer boundary in an approximately $10 \times 10$ area. The scattering problem is solved at zero bias using Kwant and the transmission probabilities between each pair of distinct leads are used to generate the full conductance matrix $G$.  

For Fig.~\ref{fig:Main:Graphene_phase_diagram}, we similarly create a disk-shaped system with $r = 60$ and attach 10 ideal leads close to its boundaries. This provides a full $10 \times 10$ conductance matrix. To evaluate the $\langle w_{\rm PD} ^{\pd} \rangle$ invariant here, we consider the last lead as grounded and work with the $9 \times 9$ $G_{\rm OBC}^{\pd}$ sub-matrix. While evaluating the trace per unit volume in Eq.~(3) and Eq.~(4), we average over 3 sites close to the middle of the chain, i.e. $\lfloor N/2 \rfloor, \lfloor N/2 \rfloor+1, \lfloor N/2 \rfloor+2$ where $N$ is the number of terminals in the OBC case and $\lfloor x \rfloor$ is the greatest integer function of x. For $N = 9$, this corresponds to the $4^{\rm th}$, $5^{\rm th}$, and $6^{\rm th}$ entries. Here, the results are averaged over $n = 400$ independent disorder realizations of the $G$ matrix. 

For Fig.~\ref{fig:Main:Graphene_Chern_number}, we construct a disk of radius $r = 15$ and the Chern number is averaged over 500 disordered configurations of the system. 

For Fig.~\ref{fig:Main:Graphene_phases_disc}, we construct a disk of radius $r = 240$ and attach 40 leads to its outer boundary. The resulting OBC conductance matrix upon grounding the last lead is $39 \times 39$. The average conductance matrix is evaluated here as $\langle G \rangle = (1/n) \sum_{j = 1}^{n} G_{j}^{\pd}$ with $n = 3000$. In panel (b) of Fig.~\ref{fig:Main:Graphene_phases_disc}, the evaluation of $\langle w_{\rm PD} ^{\pd} \rangle$ and $\langle w_{\rm PD} ^{\pd} \rangle_2$ follow the same choice of volume trace as in Fig.~\ref{fig:Main:Graphene_phase_diagram}.

\begin{figure}[tbh] 
\centering
\includegraphics[width=0.45\textwidth]{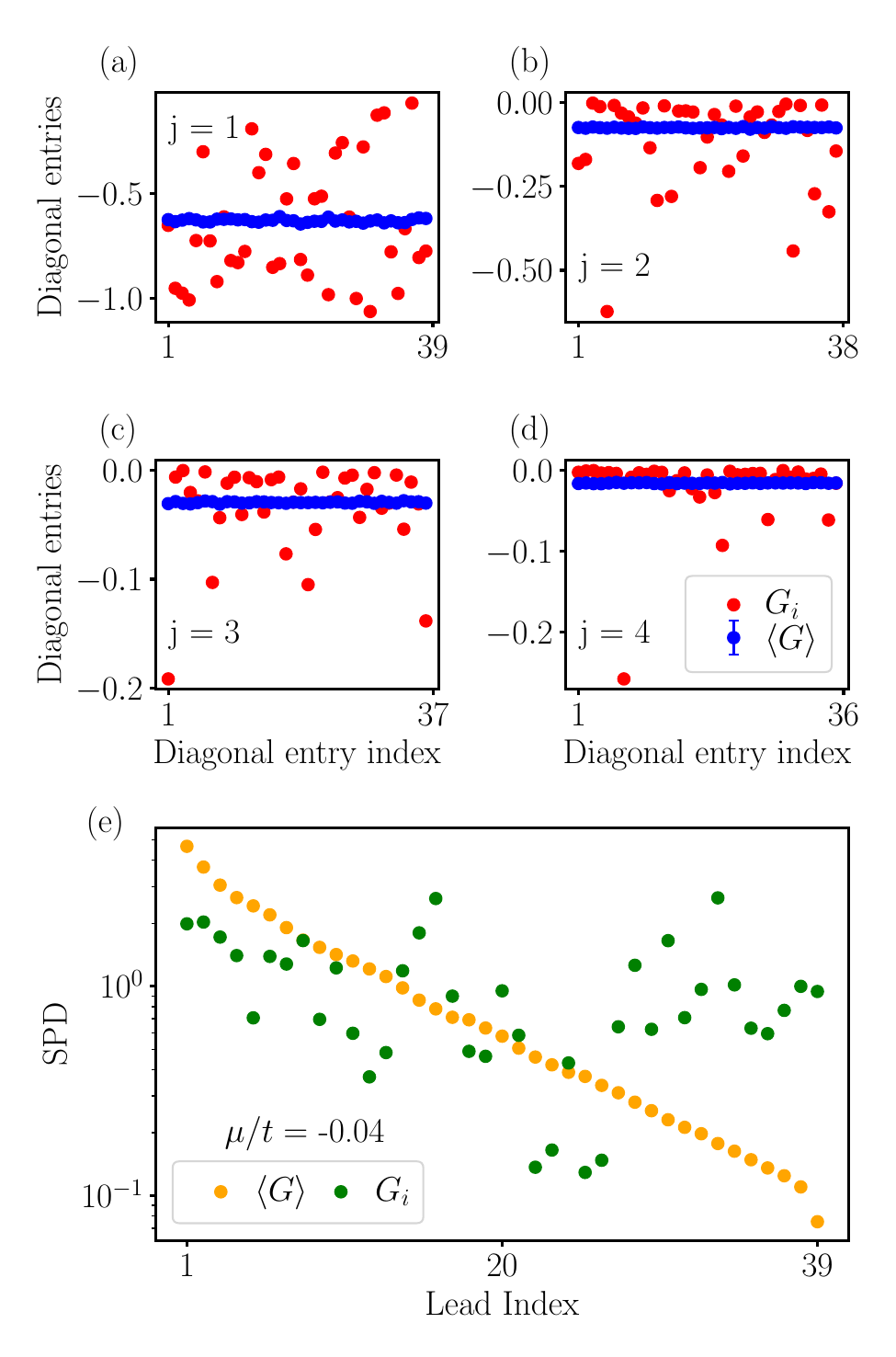}
\caption{Panels (a-d): Uniformity of the average conductance matrix. Entries of the diagonals of different orders $j$ (integer-valued) above the main diagonal ($j = 0$) of a $39 \times 39$ OBC conductance matrix (blue) and the average conductance matrix $\langle G \rangle$ (red). The average conductance matrix has uniform entries on its diagonals for diagonals of all orders $j = 1,2,3,4$ [panels (a), (b), (c), (d), respectively] whereas the entries of $G_i$ are inhomogeneous. Panel (e): SPD plotted on a logarithmic scale for a $39 \times 39$ inhomogeneous conductance matrix $G_i$ (green) and uniform average conductance matrix $\langle G \rangle$ (yellow). Due to inhomogeneities, the SPD for $G_i$ shows an absence of the non-Hermitian skin effect, i.e. it is not exponentially localized towards one end of the chain, whereas the SPD for the uniform conductance matrix exhibits it. In all the panels above, $\delta = 1.2t,$ and $\mu = -0.04t$. \label{fig:SI:uniform_conductance_matrix}}
\end{figure}

\section{Extra non-Hermitian phase}  
\label{SM:sec:wpd_scaling}

In Fig.~\ref{fig:Main:Graphene_phases_disc}, we showed the existence of an extra non-Hermitian phase for OBC conductance matrices corresponding to a system with radius $r = 240$, 40 leads, and one of the leads considered grounded.

In Fig.~\ref{fig:SI:uniform_conductance_matrix} (a-d), we tune the chemical potential $\mu$ to access this extra non-Hermitian phase and explicitly show that the average conductance matrix $\langle G \rangle$ (averaged over $n = 3000$ independent configurations) here is uniform whereas the conductance matrix $G_i^{\pd}$ for a single random onsite disorder realization $i$ of the system has inhomogeneities. Both of these $39 \times 39$ matrices are obtained from the 40 lead setup used for Fig.~\ref{fig:Main:Graphene_phases_disc}.

This extra non-Hermitian phase corresponds to a current-voltage characteristic that does not show an exponential profile of the non-Hermitian skin effect. The SPD for the inhomogeneous conductance matrix $G_i^{\pd}$ shows an absence of a piling up of its eigenstates towards one end of the chain whereas the SPD for the averaged conductance matrix shows this behavior $\langle G \rangle$, [see Fig.~\ref{fig:SI:uniform_conductance_matrix}(e)].

\section{Conductance matrix: Finite longer-range couplings} 
\label{SM:sec:Longer_range_couplings}

Consider the graphene toy model with onsite disorder $\delta$ in a finite-sized disk of radius $r= 60$ and 10 leads attached to its boundaries. Transport simulations are done for $n = 400$ independent onsite disorder realizations of the system resulting in an average conductance matrix $\langle G \rangle$.  

As the system transitions across a broadened zeroth Landau level by varying the chemical potential $\mu$, the previously topologically protected boundary modes lose their protection, leading to transport through the bulk of the system.  We demonstrate this phenomenon within the scope of our multi-terminal transport simulations, as outlined below.

\begin{figure}[tbh]
\centering
\includegraphics[width=0.45\textwidth]{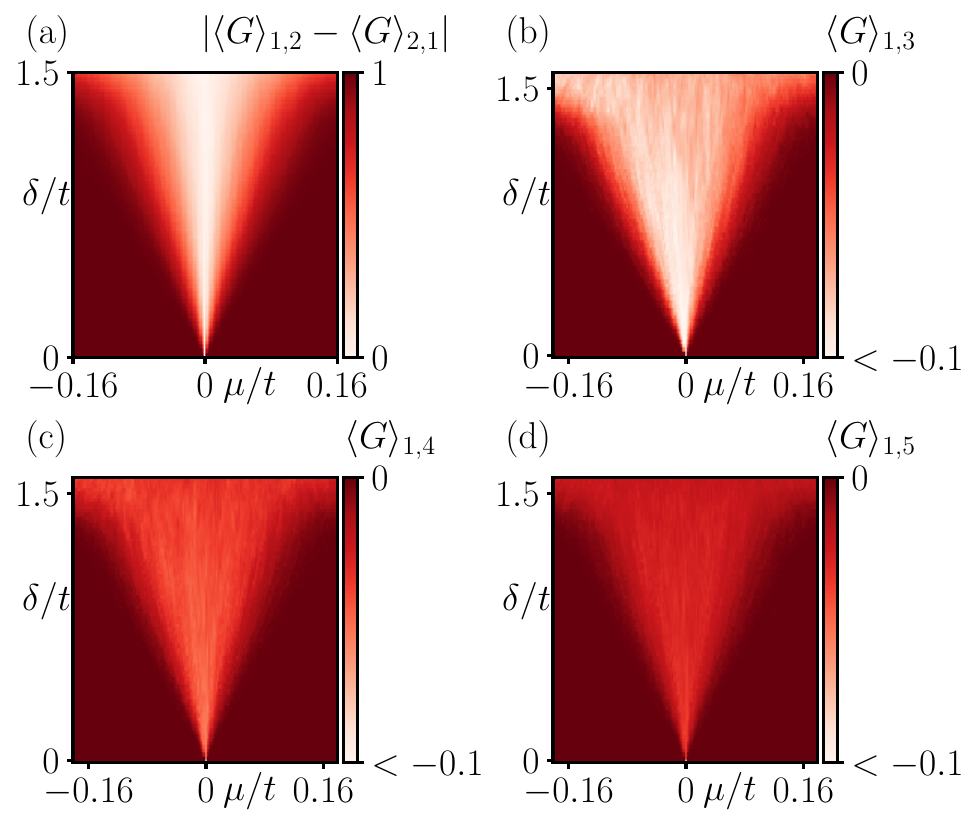}
\caption{
Average multi-terminal conductance response for a monolayer disk of graphene.  
Panel (a): Absolute difference of average conductance $|\langle G \rangle_{1,2} -\langle G \rangle_{2,1}|$ highlighting two quantum Hall phases ($|\langle G \rangle_{1,2} -\langle G \rangle_{2,1}| = 1$) separated by a broadened zeroth Landau level at the charge neutrality point $\mu/t = 0$, whose width increases with an increase in $\delta$.
Panels (b, c, d): Average conductance through the bulk of the system ($\langle G \rangle_{1,3}, \langle G \rangle_{1,4}, \langle G \rangle_{1,5}$).  
These are zero in the quantum Hall regime (edge transport) and take on finite values while transitioning between them (bulk transport). Here, $\phi = 0.18$. 
\label{fig:SI:Conductance_Phase_disorder}
}
\end{figure}

The absolute difference between $\langle G \rangle_{1,2}$ and $\langle G \rangle_{2,1}$ [see Fig.~\ref{fig:SI:Conductance_Phase_disorder}(a)] as a function of $\mu/t$ and $\delta/t$ shows the two quantum Hall phases ($|\langle G \rangle_{1,2} - \langle G \rangle_{2,1}| \approx 1$) and the transition region ($|\langle G \rangle_{1,2} - \langle G \rangle_{2,1}| \neq 1$) between them. The broadening of the zeroth Landau level here increases with the strength of $\delta/t$.

The conductance through the bulk of the system,  $\langle G \rangle_{1,3}, \langle G \rangle_{1,4}, $ and $\langle G \rangle_{1,5}$ (see Fig.~\ref{fig:SI:Conductance_Phase_disorder}(b-d)) is zero when the system resides in a quantum Hall regime, indicating no bulk transport. 
However, as the system transitions through the broadened zeroth Landau level, these entries become nonzero, reflecting the emergence of bulk transport. 
Consequently, the resulting conductance matrix exhibits finite couplings between all pairs of leads.

\begin{figure}[tbh]
\centering
\includegraphics[width=0.45\textwidth]{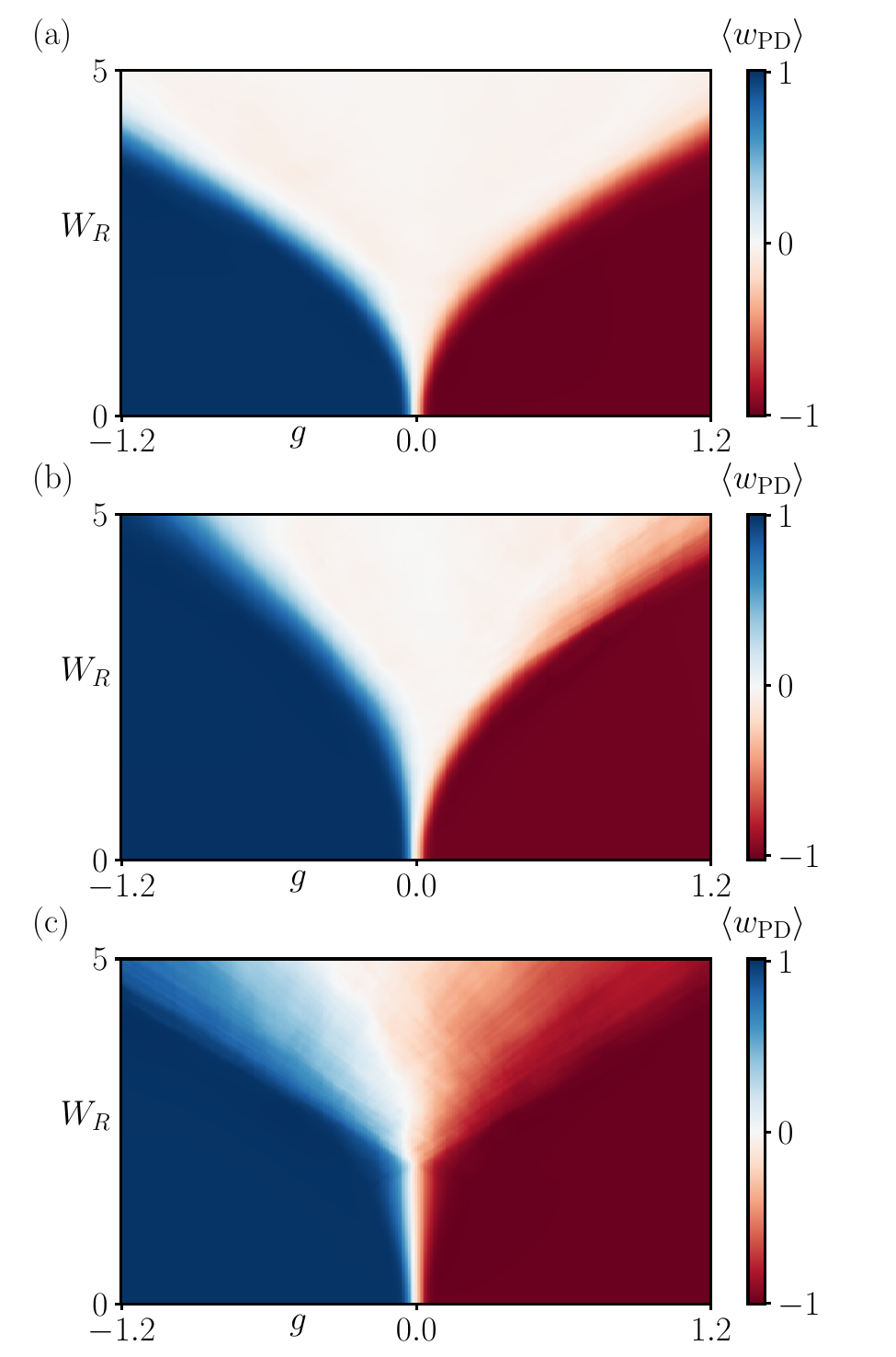}
\caption{
Panels (a,b): $\langle w_{\rm PD}^{\pd} \rangle$ in the (g, $W_{\rm R}$) phase space showing non-trivial non-Hermitian topological phases $+1$ and $-1$ separated by a trivial insulating phase when the diagonal entries of $H_{\rm HN}^{\pd}$ are constrained to satisfy the condition: sum of columns of the Hamiltonian matrix under PBC is zero; $W_{\rm R} = W_{\rm L},  W_{\rm R} = 2W_{\rm L}$ respectively. The observed asymmetry in (b) is due to the disorder strengths $W_R \neq W_L$. 
Panel (c): $\langle w_{\rm PD}^{\pd} \rangle$ in the (g, $W_{\rm R}$) phase space highlighting two non-trivial non-Hermitian topological phases ($+1$ and $-1$) when the diagonal entries of $H_{\rm HN}^{\pd}$ are manually set to zero. 
Here, we set $t_0 = 1, E_{\rm B}^{\pd}= 0$, $L = 100$.
\label{fig:SI:HN_Phase_Figure}
}
\end{figure}

\section{Trivial Phase in the Hatano-Nelson Model} 
\label{SM:sec:HN_trivial_phase}

In the main text, we showed how inhomogeneities in the entries of the conductance matrix $G$ in a multi-terminal setup of monolayer graphene could produce a trivial phase as a natural consequence of the current conservation rules.  
It is also known that a finite chain of the Hatano-Nelson model can host a trivial phase in the presence of finite onsite disorder \cite{Hughes2021}. 
Here, we leverage this knowledge and illustrate how applying constraints that mimic the current conservation rules in $G$ ($\sum_{i = 1}^N G_{ij}^{\pd} = 0$) onto the Hamiltonian matrix of a disordered Hatano-Nelson chain would lead to a trivial phase.

Consider a disordered version of the Hatano-Nelson model \cite{Hughes2021}: 
\begin{equation} \label{eq:HN_disordered}
\begin{split}
    H_{\rm HN}^{\pd} & = \sum_i J^i_{\rm right} \ket{j+1}\bra{j} + J^i_{\rm left} \ket{j}\bra{j+1},
\end{split}
\end{equation}
where $J^i_{\rm right} = (t_0+g) + W^\pd_{\rm R} w_{\rm R}^i$ and $J^i_{\rm left} = (t_0-g) + W^\pd_{\rm L} w_{\rm L}^i$ are real numbers denoting the hoppings towards the right and left, respectively. 
Here, $w_{\rm L}^i,$ and $ w_{\rm R}^i$ are random numbers independently drawn from a uniform distribution $(-0.5,0.5]$. 

The Hamiltonian matrix representing a finite chain of length $L$ for this model under open boundary conditions is given by:
\begin{equation}\label{eq:SI_HNmatrix}
H_{\rm HN}^{\pd} = 
    \begin{pmatrix}
      0 & J_{\rm left}^{1} & & & & \\
      J_{\rm right}^2 & 0 & J_{\rm left}^2 & & & \\
      & J_{\rm right}^3 & 0 & \ddots & & \\
      & & \ddots & \ddots & \ddots & \\
      & & & \ddots & 0 & J_{\rm left}^{L-1} \\
      & & & & J_{\rm right}^L & 0 \\
    \end{pmatrix},
\end{equation}
while under periodic boundary conditions, this matrix includes two additional terms ($H_{\rm HN_{1,L}} = J_{\rm right}^1$ and $H_{\rm HN_{L,1}} = J_{\rm left}^L$).

\begin{figure}[tbh]
\centering
\includegraphics[width=0.45\textwidth]{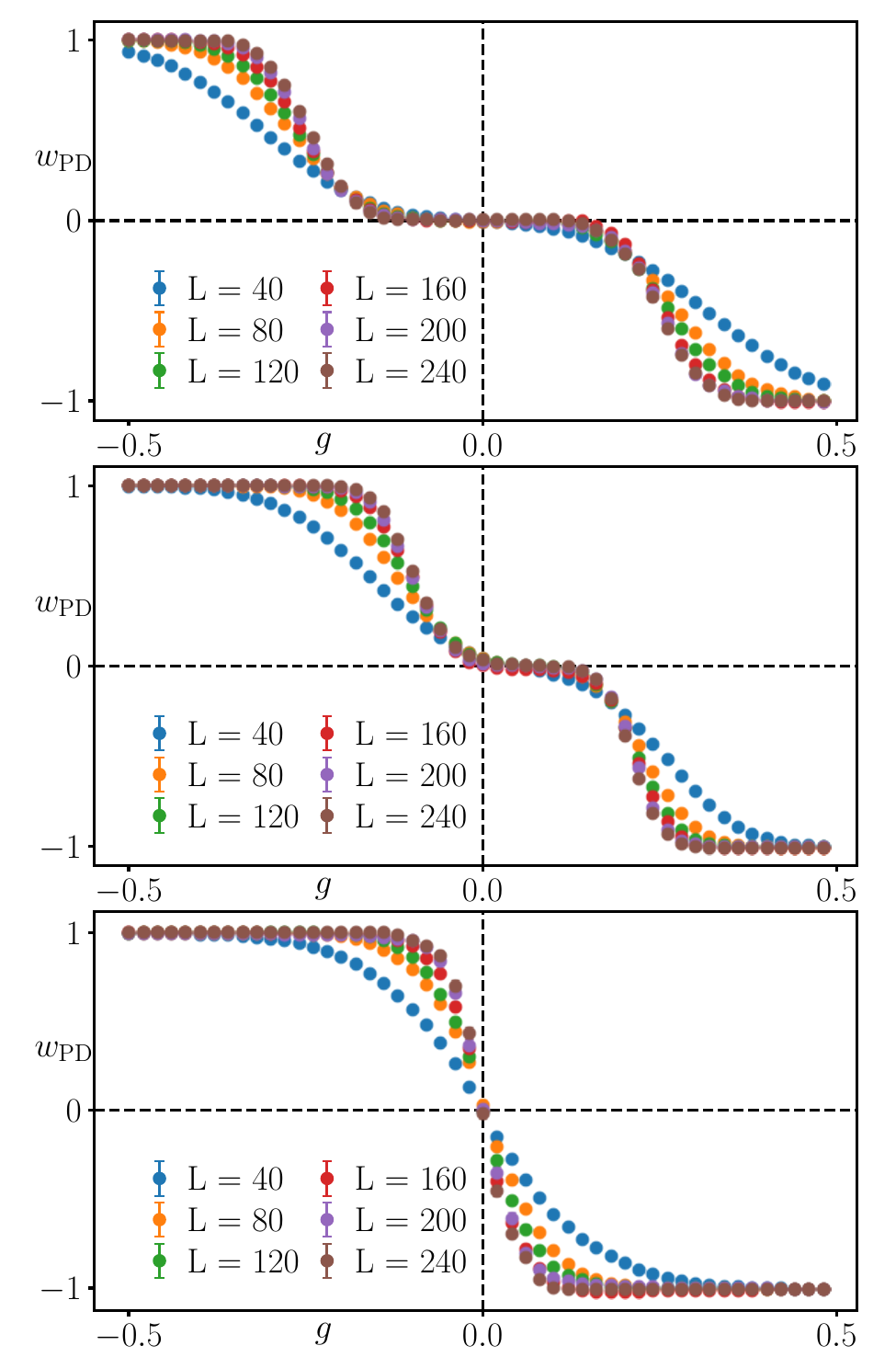}
\caption{Finite-size scaling (HN model): (a-b) $\langle w_{\rm PD}^{\pd} \rangle$ for $(W_{\rm R}, W_{\rm L}) = (2,2)$ and $(W_{\rm R}, W_{\rm L}) = (2,1)$, respectively, for different system sizes $L$ of the disordered Hatano-Nelson chain with constraints on the diagonal entries as in Fig.~\ref{fig:SI:HN_Phase_Figure} (a-b). Increasing the system size makes the phase transition much sharper between trivial and topological phases of the model. (c) $(W_{\rm R}, W_{\rm L}) = (2,2)$ and no constraint on diagonal terms: $\langle w_{\rm PD}^{\pd} \rangle$ scales with system size close to the transition and shows a direct transition between two topological phases. Here, we set $t_0 = 1, E_{\rm B}^{\pd}= 0$.
\label{fig:SI:HN_wpd_scaling}
}
\end{figure}

We now apply a constraint on the Hamiltonian matrix $H_{\rm HN}^{\pd}$ in Eq.~\eqref{eq:HN_disordered} under PBC as $\sum_{i = 1}^L H_{\rm HN_{ij}}^{\pd} = 0$, and the diagonal elements are modified accordingly. 
This leads to the disorder in the hopping terms getting translated onto the disorder in the diagonal entries of the Hamiltonian matrix in Eq.~\eqref{eq:SI_HNmatrix}. 
Similar to conductance matrices, the Hamiltonian matrix for a Hatano-Nelson chain of length $L-1$ with OBC is the $(L-1 \cross L-1)$ sub-matrix of the corresponding Hamiltonian matrix with PBC which omits the hoppings to and from the last site of the chain.

To describe the non-Hermitian topological phases for a finite size chain of this model under OBC, we evaluate the $w_{\rm PD}^{\pd}$ invariant [see Eq.~\ref{eq:wpd_invariant}] by replacing $G$ with the constrained disordered Hamiltonian $H_{\rm HN}^{\pd}$. 
Labeling the Hamiltonian matrix for a disordered configuration indexed $i$ as $H_{\rm HN_i}^{\pd}$ and setting the corresponding base point $E_{\rm B}^{\pd}= {\rm tr}( H_{\rm HN_i}^{\pd})/L$, we evaluate 
\begin{equation} \label{eq:HN_dis_wpd}
    \langle w_{\rm PD}^{\pd}\rangle = \frac{\sum_{i=1}^{n} w_{\rm PD}^{\pd}(H_{\rm HN_i}^{\pd})}{n},
\end{equation} 
where $n$ is the total number of disorder configurations. 
This disorder averaging procedure is similar to the one described in Eq.~\ref{eq:wpd-method01} for conductance matrices $G$ in the main text. 
We find that a trivial phase appears in the system alongside two non-Hermitian topological phases in the presence of disorder [see Fig.~\ref{fig:SI:HN_Phase_Figure}(a, b)].

If, on the other hand, we do not apply the constraint as above, the diagonal terms are all zero and no longer have inhomogeneities. This leads to the vanishing of the trivial phase [see Fig.~\ref{fig:SI:HN_Phase_Figure}(c)], which is expected for a disordered Hatano-Nelson chain with no on-site disorder \cite{Hughes2021}. Thus, the presence of disorder in the hopping terms combined with the physical constraint on its diagonal entries is a sufficient, but not necessary, condition responsible for the emergence of a trivial phase in this system. The trivial phase can be introduced here just as well by introducing disordered on-site terms in the original Hamiltonian Eq. (\ref{eq:HN_disordered}). \\
The above study was performed for a fixed system size $L = 100$ of the Hatano-Nelson chain. To confirm the existence of different phases in the system, we present a finite-size scaling study by taking horizontal cuts in the phase diagrams in Fig. \ref{fig:SI:HN_Phase_Figure}. As shown in Fig.~\ref{fig:SI:HN_wpd_scaling} (a,b), an increase in the system size $L$ clearly shows the $w_{PD}$ tending to a step-function marking the boundaries between the non-trivial and trivial non-Hermitian phases, whereas Fig.~\ref{fig:SI:HN_wpd_scaling} (c) shows a direct transition between the two non-trivial non-Hermitian phases. Thus, the presence of on-site disorder in the Hatano-Nelson chain introduces an extra trivial phase.

\end{appendix}
\end{document}